\documentclass[12pt,preprint]{aastex}
\usepackage{natbib}

\begin{document}

\title{Can we ignore magnetic fields in studies of PN formation, shaping and interaction with the ISM?}


\author{Noam Soker}
\affil{Department of Physics, Technion$-$Israel Institute of
Technology, Haifa 32000, Israel; soker@physics.technion.ac.il}

\begin{abstract}
Yes. \footnote{As this paper summarizes my talk, I stay as close as
possible to the presentation I gave at the Gdansk meeting.
Hence, the title is as was given by the organizers, and the abstract
is my original one.}
\end{abstract}

\maketitle


\section{ Introduction}

In a quest for better understanding the shaping mechanism of planetary
nebulae (PNs) I published many papers in the past (e.g., Soker 2004a)
all of which came to the same conclusion:
For the formation of non-spherical PNs a companion is required,
wether stellar or substellar.
This results from the basic assumption that angular momentum is a conserved
physics quantity.
As trivial as this may sound, some published models for the
shaping of PNs seem to have ignored this basic physical law.
Most recently I discussed this in a paper that was accepted by astro-ph
(Soker 2005b), but was rejected by MNRAS and the ApJ.
Both referees (a third referee in ApJ can be ignored for a
non-scientific report) were either confused, or purposely tried to mess up the
definition of magnetic shaping to save their own usage of magnetic fields.
For that, in Section 2 I repeat the classification from Soker (2005b),
which is central for any discussion to follow.
I then elaborate on several key points regarding the failure of
models for shaping PNs where dynamically important magnetic fields
are the main ingredient.

In that paper (Soker 2005b) I critically examined recent claims that the
magnetic fields are the main agent shaping PNs.
These claims are based on the detection of magnetic fields in
asymptotic giant branch (AGB) stars and in central stars of PNs.
I particular examined the energy and angular momentum
carried by magnetic fields expelled from AGB stars, as well as
other physical phenomena that accompany the presence of large
scale fields, such as those claimed in the literature.
I showed that a single star cannot supply the energy and angular
momentum if the magnetic fields have the large coherent structure
required to shape the circumstellar wind.
Therefore, the structure of non-spherical PNs cannot be attributed to
dynamically important large scale magnetic fields.
I concluded that the observed magnetic fields around evolved stars can be
understood by locally enhanced magnetic loops which can have a
secondary role in the shaping of the PN.
The primary role, I argued, rests with the presence of a companion.

My expectation was that by the time of the meeting in Gdansk, I would have
added more to what there was already in that paper, and would have presented
some new results in the meeting.
However, by the time of the meeting I found myself wasting time
dealing with stupid referee reports (see section 3).
Therefore, during my 20 minutes talk before the discussion, I had to deal mainly
with two topics: the role of the magnetic fields and the refereeing system.
As this paper summarizes my talk,\footnote{I am not good at summarizing other
people view, hence I will not write here about the discussion that followed
my talk.} I am obliged to discuss these two topics here, to the benefit
(I hope) of those who did not attend the meeting.

\section{ The role of magnetic fields in shaping PN morphologies}
\label{sec:role}
This section summarizes the main results of Soker (2005b),
as presented in the Gdansk meeting.

\subsection{ Classification}
\label{sec:role1}
Models where magnetic fields play a role in shaping the circumstellar
matter belong to one of four categories.
\begin{enumerate}
\item {\bf Strong dynamically important fields. }
Magnetic fields deflect the flow close to the stellar surface,
i.e., the magnetic pressure and/or tension are dynamically
important already on the AGB (or post-AGB) stellar surface
(e.g., Pascoli 1997; Matt et al.\ 2000; Blackman et al.\ 2001).
\item {\bf Weak dynamically important fields. }
The magnetic field is weak close to the AGB stellar surface, but
it plays a dynamical role at large distances from the star
(e.g., Chevalier \& Luo 1994; Garc\'{\i}a-Segura 1997;
Garc\'{\i}a-Segura et al.\ 1999).
\item {\bf Magnetic stellar spots and filaments.}
Local magnetic fields on the AGB stellar surface
are enhanced in cool spots and filaments.
The dust formation rate, and hence the mass loss rate, is enhanced above these
cool spots and filaments (section 2 in Soker 2004a, and references therein).
Magnetic fields never become dynamically important on a large scale.
This mechanism can operate even for very slowly rotating
AGB envelopes, and spin-up by companions even as small as
planets is enough to account for the required rotation velocity.
(This process may lead to the formation of moderate elliptical PNs
 - those with small departure from sphericity, but can't
account for lobes, jets, etc.).
\item {\bf Jet shaping.} Magnetic fields play a dominant role in the launching of jets
from accretion disks, either around stellar companions or
around the central star.
The action of the jets shapes the PN.
However, this shaping mechanism is indifferent to whether the jets
are launched by an accretion disk via magnetic (e.g., Frank et al. 2006;
Frank \& Blackman 2004) or non-magnetic processes (Soker \& Lasota 2004).
\end{enumerate}

Frank et al. (2006) presented at the Gdansk meeting the view that
\newline
{\it ``jet driven nebula''=MHD driven nebula. }
\newline
I think this is misleading, because the jet shaping in category 4 is clearly
different from the first three types of models
(MHD stands for magneto-hydrodynamics) in several aspects.
\newline
($i$) The energy source of the magnetic field is different.
In the first three categories the main
energy source is the {\it nuclear} burning in the AGB (or post-AGB) core.
It drives the convective envelope that amplifies the magnetic field.
In the 4th type (category), it is {\it gravitational} energy of the accreted gas
that is partially converted to magnetic energy.
This is a fundamental difference.
\newline
($ii$) In the first three groups the magnetic field is amplified by the AGB,
or in some models by the post-AGB, star. In the 4th group the
magnetic field is amplified by an accretion disk around the companion or
around the central post-AGB star.
\newline
($iii$) In most models of the first three categories the magnetic fields directly
influence (shape) the nebular material, as it leaves the AGB star, or at large
distances.
In the 4th category, that of jet shaping, the magnetic field influences only
a small portion of the mass that ends in the nebula (if it does at all).
Only in some models of the first category the magnetic field of the post AGB
stars shapes two fast jets which contain a small amount of mass.
\newline
($iv$) As stated above, in the fourth class of models the magnetic field is
not a necessary ingredient; it is enough that two jets (or a collimated fast wind
[CFW]) are launched, whether by magnetic fields, thermal pressure, or other mechanisms.
\newline
($v$) As I have shown in several papers (Harpaz \& Soker 1994; Soker 1996;
Soker \& Zoabi 2002), in the first three classes the AGB or post-AG B stars
must be spun-up by a stellar (or substellar in the third class) companion.
In the 4th class the accretion disk is formed when the transferred mass has
high enough specific angular momentum resulting from the orbital motion.
\newline
($vi$) As stated, models for launching jets without magnetic fields exist.
But even if magnetic fields play a role in launching jets from accretion
disks (as most researchers think), so does the viscosity in the disk.
Still, we don't term the jet shaping viscous shaping.

For our purpose, the main question is what ingredients
are required to form elliptical and bipolar PNs.
The question of how accretion disks form jets is common to many fields
in astronomy, and it will not advance our understanding of the PN jet-shaping
models. The main question for PNs jet-shaping is: What are the conditions for
the formation of accretion disks in PN progenitors.

\subsection{ General considerations}
The main questions regarding shaping of bipolar and elliptical PNs concern the
evolutionary track which will bring the progenitor, whether a single star
or a binary system, to the conditions leading to axisymmetrical mass loss.
Examples for such questions are:
\begin{itemize}
\item  What is the evolutionary rout to form accretion disks around a companion?
\item What are the circumstances by which a stellar companion
spins-up the AGB envelope via tidal interaction?
\item What are the circumstances by which a companion, stellar or substellar,
spins-up the AGB envelope via common envelope evolution?
\item Under what conditions a stellar companion outside the envelope deflects
the AGB wind to form a dense equatorial flow?
\item What happen to a low mass companion spiraling all the way to the AGB
core and then destroyed there?
\item Which morphological features are formed by jets?
\item Are all non-spherical PNs shaped by jets?
\item Can a single star acquire any of these properties without the assistance
of a companion?
\end{itemize}
My view and answer to these questions are summarized in Soker (2004a), where
a list of papers can be found, while a general review of PN shaping,
updated to the year 2002, can be found in Balick \& Frank (2002).

Some papers avoid these questions, skip evolutionary considerations, and {\it assume}
the asymmetry from the start.
I don't think these papers advance our understanding of the shaping of planetary
nebulae in any significant way. 
An example is the paper by Matt et al.\ (2004), who consider magnetic field
in the core-envelope boundary.
A similar idea was treated already by Pascoli (1997);
in a previous paper (sec. 2.2. in Soker 1998) I discussed the
problems I see in the model of Pascoli.
Matt et al. (2004) start with a fast rotating core with a very
strong magnetic dipole field.
Namely, the asymmetry is already there, and it is assumed to occur at a specific
time, without any evolutionary considerations. It is not clear how AGB or post-AGB
stars can reach such a state.
It is not enough to state that a companion will do this. Other effects of
the companion must be considered.

One of the processes that many MHD models skip is angular momentum evolution
of the AGB star.
In several previous papers (e.g., Harpaz \& Soker 1994; Soker 1996;
Soker \& Zoabi 2002) it was shown that the {\it mass} lost by the AGB star will
substantially slow down the AGB envelope as the wind caries angular momentum.
This by itself shows that any MHD model must have a companion
to spin-up the envelope. But then magnetic effects cannot be considered
by themselves; other effects of the companion must be considered.
In addition to the angular momentum evolution, there are other problems
with magnetic models which are summarized in these papers.
This conclusion holds not only to PNs.
In Soker (2004b) I have discussed the case of the bipolar nebula around
the Massive binary system $\eta$ Carinae, and concluded that it could not
have been formed by a single star.
Some processes related to the shaping of the $\eta$ Carinae nebula and
other bipolar nebulae are discussed in Soker (2005a), where I also criticized
the $\eta$ Carinae magnetic shaping model of Matt \& Balick (2004).

In my presentation at the meeting and in my recent paper (Soker 2005b) I showed
that not only the mass in the wind carries angular momentum that
substantially spins-down the star, but the {\it magnetic field itself} does so
(with the mass frozen to the magnetic fields at large distances from the star).
This shows that most magnetic shaping models are not even self-consistent.
I emphasize again:
(1) This does not mean that magnetic fields don't play any role; it just shows
that any process involving dynamically important magnetic fields must
be incorporated into a binary model, and cannot stand by itself.
(2) Shaping by jets that are launched by accretion disks are not considered to
be magnetic models.

A few words on the interacting winds models, where a fast wind from the central
star of the PN interacts with the previously ejected asymmetrical slow AGB wind
(also called GISW model), are in place here.
This process by itself is not a model for asymmetrical PNs, as it starts with
an already asymmetrical slow wind; it is only an additional process in many models.
The fast wind, as well as ionization fronts (e.g., Mellema 1995;
Soker 2000; Perinotto et al. 2004), only changes the already asymmetrical mass
distribution in the slow wind.
As was noticed from the first days of 2-dimensional numerical simulations
of the winds interaction process in PNs (Soker \& Livio 1989), a binary companion
is the most likely explanation for the asymmetrical slow wind.
It was also noticed already in 1989 (Soker 1990\footnote{In an ugly way two referees
rejected that paper from ApJ in 1989. One of them later obtained the same result,
namely, that ansae cannot be formed by the wind interaction process.}) that the winds
interaction cannot explain {\it ansae}$-$two small dense blobs along the symmetry
axis of elliptical PNs (termed also FLIERS).
I suggested then that two jets are blown during the post-AGB stage.
Therefore, contrary to some claims, the knowledge that the interacting wind model cannot
account for all features in PNs did not come only with the high quality
HST images in 1995-6; it was there several years earlier.
The idea of jet shaping in a binary system was put forward even earlier by
Morris (1987).

\subsection{ Angular Momentum}
I will not repeat the calculations presented in Soker (2005b, astro-ph/0501647),
as the interested reader has a free access to that paper.
I will only mention the main points.

Basically, magnetic fields carry angular momentum as they are bent by the rotating
star and the material frozen to the field lines at large distances from the star.
This is connected to the force due to tension not having a pure radial component,
as the required magnetic fields in magnetic shaping models.
In Soker (2005b) I demonstrated that large magnetic fields will carry
angular momentum away from the stellar envelope much faster than mass is lost
by the wind. This has the effect of slowing down the star on short time
scales.
Since the magnetic field is powered by rotation, this argument disproves that
such a strong large scale magnetic field is present in the star.

The expressions for the spin-down rate due to the magnetized wind
were applied to objects claimed to have strong magnetic fields around
them (e.g., Miranda et al.\ 2001; Vlemmings et al.\ 2002, 2005; Diamond \& Kemball 2003;
Bains et al.\ 2003, 2004a,b; Szymczak \& Gerard 2004, 2005;
Jordan et al.\ 2005), and the conclusion by some authors that these
magnetic fields have had a role in shaping their PNs was criticized.
By building on the arguments of Soker \& Zoabi (2002) and Soker (2004a) I
exposeed additional reasons why these conclusions are unjustified.

In order to influence the large scale structure of the circumstellar
matter (the wind), the magnetic field must have a large scale structure.
However, I showed that if the observed magnetic fields have a large scale
structure, then they contain more angular momentum and energy
than a single star can supply.

I also discussed that any claim that the strong observed magnetic
fields demonstrate that the PN was shaped by them, must consider
several physical characteristics of the system that can make
global fields implausible PN shaping agents.
For example, as mentioned in Soker (2005b), in some cases the surface magnetic
pressure is larger than the thermal pressure on the entire surface
(not only in small areas as in solar spots); this does not make
sense as the photosphere of the star would be unstable.
As discussed there (Soker 2005b), the models for large
scale magnetic shaping must consider the stress of the magnetic field.
When this is done, some contradictions between observation
and the expected behavior of the field appear.

By demonstrating that the magnetic fields observed cannot play a global
role in shaping the PN, I am not denying that magnetic fields can exist
in AGB stars and PNs. These observed fields
can be attributed to local ejection events, similar to the magnetic clouds
in the solar wind (Soker \& Kastner 2003).
Such locally-enhanced magnetic fields can be formed by a dynamo mechanism
rooted mainly in the vigorous convective envelope of AGB stars.
For this mechanism to operate the AGB star does not need to rotate fast;
slow rotation is sufficient and can be achieved even by an orbiting
planet spinning up the envelope.
Local magnetic fields can influence the wind geometry by
facilitating dust formation.
Locally enhanced magnetic fields on the AGB stellar surface can
lead to the formation of cool
spots and cool filaments, as is the case in the Sun.
Dust formation rate, hence mass loss rate, is enhanced above these cool
spots and within the cool filaments (Soker \& Zoabi 2002).
If spots are concentrated near the equator, then mass loss rate is
higher in that direction.
This process might lead to the formation of moderately elliptical PNs
(those with small departure from sphericity), but
cannot account for lobes, jets, etc. Hence, these local magnetic fields
can play a role in the PN morphology but this is a secondary role
to that exercised by companions to the AGB star.

Ignoring the effects of binary companions in the shaping of PNs and related
nebulae, will lead to erroneous conclusions. For instance,
Jordan et al.\ (2005) argue that their discovery of magnetic
fields on the surface of four central stars supports the
hypothesis of magnetic shaping.
In the case of systems like EGB 5 (PN G211.9+22.6), one of the four systems
they discuss, the known WD companion (Karl et al. 2003)
is expected to influence the mass loss geometry more than any
possible magnetic field in AGB stars.

As I have stated and demonstrated in previous papers
(Soker 2004a, and references there in) the effect of companions
to the AGB stars, even as small as brown dwarfs and planets,
can give rise to a host of physical phenomena, resulting in shaping
of the AGB mass-loss and subsequent wind into the morphologies observed.
Circular PNs don't need any companion to shape the wind, and can be formed
by single stars.

\section{ On the refereeing system}
\label{sec:referee}
\subsection{ Examples}
\label{sec:referee1}

To demonstrate the type of referee comments I have to deal with, I bring two examples
from referee reports on the paper that was described in the previous section (Soker 2005b).

One of the comments made by the MNRAS referee to Soker (2005b):
\newline
{\it ``A related issue is that the author does not discuss the results of Bujarrabal et al. 2001,
which show observationally that the directed momentum from radiation is not enough to power
the collimated flows of protoplanetary nebulae. It is quite plausible that a binary is important
in supplying the needed angular momentum, but the possibly fundamental intermediary role of the
magnetic field in launching a jet is still fully consistent with the need for a binary. ''}

My minor comment is related to the confusion of what magnetic shaping means.
As discussed above, the main issue is the launching of jets via accretion disks.
Magnetic field are thought by many to play a significant role in that, but this is
not magnetic shaping!
The magnetic field may, or may not, collimate the jets, but the nebula
itself is shaped by the jets.
My major comment to the referee point is that in Soker (2002) a subsection
(section 2.1 there) is devoted to explain the results of Bujarrabal et al.
The first sentence of section 2.1 in Soker (2002) reads:
\newline
{\it ``The purpose of this section is to show that the proto-bipolar PN
OH 231 can be naturally explained by a binary model, as was suggested
already by Cohen et al. (1985), despite claims of Alcolea et al. (2001)
and Bujarrabal et al. (2001) that there is no satisfactory theory to
account for this and similar bipolar PNs.''}
\newline
Four years after I specifically accounted for the results of Bujarrabal
et al. (2001) in the frame of the binary model for PN shaping, the referee
rejects my paper based on this and similar unfair or nonscientific comments.

The reason I write that the referee {\it rejects}, and not that the referee
{\it recommends rejection}, is that in the cases I know, the editor always
accepts the referee verdict, as stupid as it may be\footnote{In one case this
was not true. A referee recommended accepting one of my papers for publication
[I know this both from the referee himself and the editor], but the
editor wrote me that the referee is not an expert in the field
[this is the referee they chose!], so they asked another referee who rejected
the paper. }

The second example is from the ApJ referee who writes in one of his comments:
\newline
{\it The author seems to have several confusing statements about magnetic and
angular momentum transport. First, ANY mass loss mechanism in a rotating and
magnetized star also implies the corresponding angular momentum and magnetic losses.
However, this does NOT imply that a net torque is operative inside the star,
not that the internal magnetic energy density decreases in time.
Hence, the "conlusion" that a companion is needed to spin-up the stellar envelope
is at best, a clear missunderstanding.}
\newline
As I wrote the editor, this is a stupid comment.
The MHD models criticized in Soker (2005b) require the envelope to rotate fast.
The loss of angular momentum will slow down the outer layer of the star,
and fresh angular momentum is required to be supplied from the deep
envelope if these models are to work at all.
This, as I have shown in several previous papers
(Harpaz \& Soker 1994; Soker 1996; Soker \& Zoabi 2002),
will very rapidly slow down the envelope, unless a companion is present.
How come the referee, who is at best an ignorant referee, does not get the main
points of the paper but allows himself to write a report with so much confidence?

\subsection{ A suggestion to abolish journals and to change the refereeing system}
\label{sec:referee2}

The refereeing system suffers from two major flaws.
\newline
(1) Many referees exploit their anonymity, don't check their claims in the report,
and allow themselves to make ignorant remarks.
As I wrote previously (Soker 2004a, sec. 4.1),
the present situation, where someone can present his or her scientific
view in a referee report without the need to stand behind this view
(because it is not going to be published anywhere) is scientifically intolerable.
In many cases the referee comments are not even scientific.
\newline
(2) In many cases editors don't do the job expected from them.
They automatically accept the referee recommendation without applying
any scientific judgement.
I criticized this editorial behavior in a letter to the chief editor of
one of our journals (I received no reply). The last sentence of my letter was:
``I expect the editorial process to {\it shrink referees to their
natural size}, and limit their power.''

 There are several alternatives which are much better than the present situation.
I expect most readers' view to be that the present refereeing system is not
perfect, but it is working. Well, I think it is not working.
I think that with the very successful astro-ph system (the arXiv)
the journals become unnecessary, and they are becoming a `counting machine' of
refereed papers for the sole purpose of promoting and hiring people.

As one of the possible alternatives, I suggest the following.
Journals in astronomy will be closed (say within 3 years, allowing each
researcher to contribute her/his last paper to a {\it refereed journal}).
The system will be astro-ph (or something similar), where researchers
can post their papers.
The new ingredient I suggest is that there will be an official refereeing rubric,
working as follows.
The authors of a paper will send their paper to a referee or several referees
they themselves choose.
As today, authors will be able also to post their papers without any refereeing,
and either leave the paper like that, or later ask for a referee.
The refereeing process will be between the authors and the referees.
After the authors are satisfied, and they have the recommendation of one
or more referees (as they choose), they will add a line to astro-ph
where the name(s) of the referee(s) appears (the referee will get a note
to acknowledge that, to prevent misunderstandings).

Since the name of the referee will appear together with the paper, the referee
will have to take much-much greater responsibility when reviewing papers.
In some cases the referee may want to approve a paper beside one point.
Then the authors can let the referee write a page or two at the end of their paper,
raising his/her objection. This will add a lot to the scientific discussion.

I note the following:
\begin{itemize}
\item Authors will have no interest in picking a "friendly" referee so that
they can count another refereed paper.
The readers will know who the referee is, which is now sharing part
of the responsibility.
For example, readers will notice when there is a small closed
group who endorse each other's papers.
The referee, with his name openly known, will have no interest in endorsing
what he/she considers as a bad paper.
\item Some refinements can be made.
For example, only people which have already, say, 20 refereed papers will be
allowed to sign as referees. This will prevent cases (which occur in
the present system) where students and young postdocs get to referee papers
in fields they know almost nothing about.
\item In hiring and promoting people, these papers can be treated as papers in
refereed journals today.
\item This system will save the page charge, which adds up to lots of money
for many people.
\item All papers appearing in astronomy will be available to all researchers.
Presently, some journals are too expensive, and practically to read a paper
one goes to astro-ph.
\item The working atmosphere will improve.
For example, in going to meetings we will not be suspicious, searching for
the person who rejected our paper a year ago, etc.
\item We will free tens of good researchers who now spend part of their
time as editors (some of them bad editors), to go back to full research.
\end{itemize}





{\bf Acknowledgments.}
I thank the organizing committee for allowing me to present my
view, something I was prevented from doing in some journals.
I also thank the organizers for allocating plenty of time for discussion
during this meeting.



\bibliographystyle{aipproc}   


\begin{thebibliography}{}

\bibitem{A} Alcolea, J., Bujarrabal, V., Sanchez Contreras, C., Neri, R.,
  \& Zweigle, J. 2001, A\&A, 373, 932

\bibitem{B1} Bains, I., Gledhill, T. M., Richards, A. M. S., \&
   Yates, J. A. 2004a, in Asymmetrical Planetary Nebulae III:
   Winds, Structure and the Thunderbird,  eds. M. Meixner,
   J. H. Kastner, B. Balick, \& N. Soker, ASP Conf. Series, 313,
   (ASP, San Francisco), 186

\bibitem{B2} Bains, I., Gledhill, T. M., Yates, J. A., \&
    Richards, A. M. S.  2003, MNRAS, 338, 287

\bibitem{B3} Bains, I., Richards, A. M. S., Gledhill, T. M., \&
    Yates, J. A.  2004b, MNRAS, 354, 529

\bibitem{Ba} Balick, B., \& Frank, A.\ 2002, ARA\&A, 40, 439

\bibitem{bl} Blackman, E. G., Frank, A., Markiel, J. A.,
  Thomas, J. H., \& Van Horn, H. M. 2001, Nature, 409, 485

\bibitem{bu} Bujarrabal, V., Castro-Carrizo, A., Alcolea, J., \&
   Sanchez Contreras, C. 2001, A\&A, 377, 868

\bibitem{ch} Chevalier, R. A., \& Luo, D. 1994, ApJ, 421, 225

\bibitem{co} Cohen, M.; Dopita, M. A.; Schwartz, R. D.;
    Tielens, A. G. G. M. 1985, ApJ, 297, 702

\bibitem{di} Diamond, P. J., \& Kemball, A. J., 2003, ApJ, 599, 1372

\bibitem{fr1} Frank, A. et al. 2006, these proceedings

\bibitem{fr2} Frank, A. \& Blackman E. G. 2004, 614, 737

\bibitem{ga1} Garc\'{\i}a-Segura, G. 1997, ApJ, 489, L189

\bibitem{ga2} Garc\'{\i}a-Segura, G., Langer, N., Rozyczka, M., \&
   Franco, J. 1999, ApJ, 517, 767

\bibitem{hs} Harpaz, A., \&  Soker, N. 1994, MNRAS, 270, 734

\bibitem{jo} Jordan, S., Werner, K., \& O'Toole, S. J.
        2005, A\&A, 432, 273

\bibitem{ka}  Karl, C., Napiwotzki, R., Heber, U., Lisker, T.,
   Nelemans, G., Christlieb, N., \ Reimers, D. 2003,
   in White Dwarfs, eds. D. de Martino, R. Silvotti, J.-E. Solheim,
   R. Kalytis. , NATO Science Series II, Vol. 105, 43

\bibitem{mab}  Matt, S., \& Balick, B. 2004, ApJ, 615, 921

\bibitem{mat} Matt, S., Balick, B., Winglee, R., \& Goodson, A.
2000, ApJ, 545, 965

\bibitem{maf}  Matt, S., Frank, A., \& Blackman, E. 2004,
   in Asymmetrical Planetary Nebulae III: Winds, Structure and the Thunderbird,
   eds. M. Meixner, J. H. Kastner, B. Balick, \& N. Soker, ASP Conf. Series, 313,
   (ASP, San Francisco), p. 449 (astro-ph/0308548)

\bibitem{m} Mellema, G. 1995, MNRAS, 277, 173

\bibitem{mi} Miranda, L. F., Gomez, Y., Anglada, G., \&
 Torrelles, J. M. 2001, Nature, 414, 284

\bibitem{mo} Morris, M. 1987, PASP, 99, 1115

\bibitem{pas} Pascoli G. 1997, ApJ, 489, 946

\bibitem{per} Perinotto, M., Sch\"onberner, D., Steffen, M., Calonaci, C.
      2004, A\&A, 414, 993

\bibitem{so90} Soker, N.\ 1990, AJ, 99, 1869

\bibitem{so96} Soker, N.\ 1996, ApJ, 469, 734

\bibitem{so98} Soker, N.\ 1998, MNRAS, 299, 1242

\bibitem{so00} Soker, N. 2000, MNRAS, 318, 1017

\bibitem{so02} Soker, N. 2002, MNRAS, 330, 481

\bibitem{so04a} Soker, N. 2004a, in Asymmetrical Planetary Nebulae
  III: Winds, Structure and the Thunderbird, eds. M. Meixner,
  J. H. Kastner, B. Balick, \& N. Soker, ASP Conf. Series, 313,
  (ASP, San Francisco), p. 562 (extended version on astro-ph/0309228)

\bibitem{so04b} Soker, N.\ 2004b, ApJ, 612, 1060

\bibitem{so05a} Soker, N.\ 2005a, ApJ, 619, 1064

\bibitem{so05b} Soker, N.\ 2005b, accepted by astro-ph/0501647

\bibitem{sok} Soker, N.\, \& Kastner, J. H.  2003, ApJ, 592, 498

\bibitem{sol} Soker, N. \& Lasota, J.-P. 2004, A\&A, 422, 1039

\bibitem{sol2} Soker, N. \& Livio, M. 1989, ApJ, 339, 268

\bibitem{soz} Soker, N. \& Zoabi, E. 2002, MNRAS, 329, 204

\bibitem{sz1}   Szymczak, M., \& Gerard, E. 2004, A\&A, 423, 209

\bibitem{sz2}   Szymczak, M., \& Gerard, E. 2005, preprint

\bibitem{vl1} Vlemmings, W. H. T., Diamond, P. J., \& van Langevelde H. J.
   2002,  A\&A, 394, 589

\bibitem{vl2}   Vlemmings, W. H. T., van Langevelde H. J., \& Diamond, P. J.,
  2005,  A\&A, 434, 1029

\end{thebibliography}



\end{document}